\documentclass{article}
\usepackage{graphicx}

\begin{document}

\begin{flushleft}
Report to the 16th International Spin Physics Symposium \\
October 10--16, 2004, ICTP, Trieste, Italy
\end{flushleft}
\vspace*{1.0cm}
\begin{center}
{\Large\bf Spin transfer to $\Lambda_{c}^{+}$ hyperons in polarized
  proton collisions at RHIC}
\end{center}

\begin{center}
V.~L.~RYKOV and K.~SUDOH
\end{center}

\begin{center}
{\em\small RIKEN, 2-1 Hirosawa, Wako, Saitama 351-0198, Japan \\
E-mails: rykov@riken.jp, sudou@rarfaxp.riken.jp}
\end{center}

%\maketitle

%\abstracts{
The analysis\cite{ohkuma:1} of helicity transfer to $\Lambda_{c}^{+}$
in polarized proton collisions is extended to the proton helicity
correlations with the $\Lambda_{c}^{+}$ transverse polarization in the
production plane (parameter $D_{LS}$). The available spin transfer
observables for the collisions of {\em two} longitudinally polarized
protons are evaluated. It is shown that, in the central region at
$\Lambda_{c}^{+}$ transverse momenta of a few GeV/c, $D_{LS}$
parameters are of about the same size as the helicity-to-helicity
correlations. The methodical issue of using spin transfers for
cross-checks of systematic errors in cross-section $A_{LL}$
measurements at polarized proton colliders is also briefly discussed.
%}

\section{Introduction}
Spin transfers to inclusive strange and charmed hyperons in polarized
proton collisions have been recently proposed\cite{ohkuma:1,xu:1} as a
probe for the polarized gluon distribution $\Delta G/G$ of
proton. Compared to the usually considered for this purpose
cross-section asymmetry
$A_{LL}=\frac{\sigma^{++}-\sigma^{+-}}{\sigma^{++}+\sigma^{+-}}$,
where $\sigma^{++}$ and $\sigma^{+-}$ are the cross sections for same
and opposite helicities of colliding protons, spin transfers are
linear with $\Delta G/G$ while $A_{LL}\propto (\Delta G/G)^{2}$. This
means that spin transfers might become more sensitive probes for
polarized gluon distribution if $\Delta G/G$ appeared to be small. The
other difference is that, unlike $A_{LL}$, spin transfer measurements
generally do not require monitoring the relative luminosity of
collisions with different polarizations of initial protons. Such
monitoring is not a simple task at a proton collider with
longitudinally polarized beams and is always considered as a potential
source of systematic errors. And, in general, measuring a number of
sensitive characteristics rather than one and comparing them to the
predictions of theoretical models could serve as a good consistency
check of the model's assumption.

In the papers\cite{ohkuma:1}, the measurements of helicity-to-helicity
transfer parameter $D_{LL}$ in gluon fusion dominant $\Lambda_{c}^{+}$
production at RHIC with polarized protons have been proposed and
studied\footnote{The notation $A_{LL}$ has been used for $D_{LL}$ in
  Refs\cite{ohkuma:1}.}. In this report, we extended this analysis to
the proton helicity correlations with the $\Lambda_{c}^{+}$ transverse
polarization in the production plane (parameter $D_{LS}$)\footnote{$L$
  and $S$ axes here correspond to $Z$ and $X$ in the notations of
  book\cite{leader:1}.}. The $D_{LS}$ is also expected to be nonzero
at $\Lambda_{c}^{+}$ transverse momenta ($P_{T}$) of a few GeV/c due
to the large $c$-quark mass. Moreover, for each spin transfer, $LL$
and $LS$, we evaluated two more observables: $D_{L\Pi}^{++}$ and
$D_{L\Pi}^{+-}$, $\Pi=L,S$, which will be measured at RHIC in
collisions of {\em two} polarized protons of the same and opposite
helicities:
\begin{eqnarray}
D_{L\Pi}^{++} & = & \frac{\sigma_{L\Pi}^{++;+}-\sigma_{L\Pi}^{++;-}-
  \sigma_{L\Pi}^{--;+}+\sigma_{L\Pi}^{--;-}}
{\sigma^{++;+}+\sigma^{++;-}+
  \sigma_{L\Pi}^{--;+}+\sigma_{L\Pi}^{--;-}}\,, \nonumber \\
D_{L\Pi}^{+-} & = & \frac{\sigma_{L\Pi}^{+-;+}-\sigma_{L\Pi}^{+-;-}-
  \sigma_{L\Pi}^{-+;+}+\sigma_{L\Pi}^{-+;-}}
{\sigma_{L\Pi}^{+-;+}+\sigma_{L\Pi}^{+-;-}+
  \sigma_{L\Pi}^{-+;+}+\sigma_{L\Pi}^{-+;-}}\,,\;\;\; \Pi=L,S
\label{eq:d_pol}
\end{eqnarray}
In Eqs.~(\ref{eq:d_pol}), $\sigma_{LS}^{--;+}$, for example, is for the
production cross-section of $\Lambda_{c}^{+}$ with the polarization
``+1'' along the $S$-axis in the collisions of two proton beams, both
of the {\em negative} helicity equal to ``-1''.

Parameters $D_{L\Pi}$ for collisions of polarized protons at
unpolarized are the weighted with $A_{LL}$ averages of $D_{L\Pi}^{++}$
and $D_{L\Pi}^{+-}$:
\begin{equation}
D_{L\Pi} = \frac{1}{2}
[D_{L\Pi}^{++}(1+A_{LL})+D_{L\Pi}^{+-}(1-A_{LL})] 
\label{eq:d_unpol}
\end{equation}
In turn, if all three $D$'s for the same final spin component were
measured, then $A_{LL}$ can be derived, using
Eq.~(\ref{eq:d_unpol}). As it mentioned above, the $A_{LL}$ determined
this way would potentially be free from systematics due to monitoring
the relative luminosity of collisions with different beam
polarizations. With this  approach, the statistical error $\delta
A_{LL}\approx\frac{2\sqrt{6}}{\alpha P|D_{L\Pi}|\sqrt{N}}$, where
$\alpha$ is the hyperon decay asymmetry parameters; $P$ is the beam
polarization; $N$ is the combined statistics in 3 measurements. This
error would usually be noticeably larger than of ``direct'' $A_{LL}$
measurements. However, if the systematic rather than statistic is an
issue, then using spin transfers in high event rate processes, along
with Eq.~(\ref{eq:d_unpol}), could be an option.

\section{Numerical results and discussion}
%%%%%%%%%%%%%%%%%%%
\begin{figure}[htb]
  \centerline{
    \includegraphics*[width=1.0\textwidth,clip]{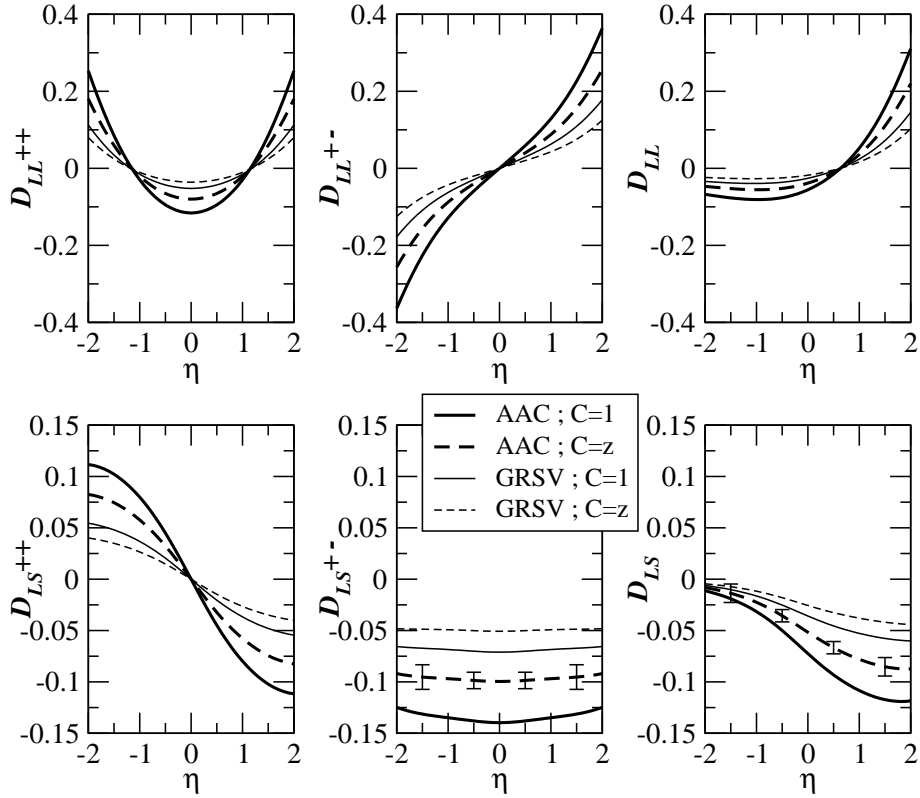}
  }
  \caption[]{
    $\eta$-dependences of spin transfer parameters for inclusive
    $\Lambda_{c}^{+}$ production in polarized proton collisions
    at $\sqrt{S}$= 200~GeV. The leading order predictions for {\em
    AAC}\,\cite{goto:1} and {\em GRSV}\,\cite{gluck:1} polarized gluon
    distributions are compared. Each error bar is for the integrated
    statistics within a pseudo-rapidity interval of $\Delta\eta =
    1$. See text for other details.
  }
\label{fig:all_d}
\end{figure}
%%%%%%%%%%%%%%%%%%%
The leading order calculations for pseudo-rapidity dependences of 6
spin transfer parameters, averaged over $P_{T}$ interval from 2 to
5~GeV/c, are shown in Fig.~\ref{fig:all_d}. These results have been
obtained, using the same assumptions as in the
analyses\cite{ohkuma:1}. Only the dominant partonic subprocess of
gluon fusion, $gg\rightarrow c\overline{c}$, was taken into
account. The same spin dependent fragmentation function
$\Delta{\mathcal D}(z) = {\mathcal C}(z)\cdot{\mathcal D}(z)$ were
used for both the longitudinal and transverse spin transfers from
$c$-quark to $\Lambda_{c}^{+}$, where ${\mathcal D}(z)$ is the
``unpolarized'' quark fragmentation function. For the ${\mathcal
  C}(z)$, two options are compared: ${\mathcal C}(z)=1$ and ${\mathcal
  C}(z)=z$. The shown statistical errors are for the integrated
luminosity of 320~pb$^{-1}$ and beam polarization of 70\%, assuming
that the decay chain
$\Lambda_{c}^{+}\rightarrow\Lambda^{0}\pi^{+}\rightarrow
p\pi^{-}\pi^{+}$ is to be used for measuring the $\Lambda_{c}^{+}$
polarization, with the detection efficiency at $\sim$10\%.

In the central region of $|\eta|<1$, $D_{LL}$'s and $D_{LS}$'s are of
about the same size in the range of $\sim$5--15\%. As $\eta$
increases, all $D_{LL}$'s grow up to $\sim$20--30\% at $\eta\sim 2$
for the {\em AAC} parameterization\cite{goto:1}, while $D_{LS}^{+-}$
stays almost flat. The achievable statistical errors of about 1\% are
small enough to clearly separate predictions for the shown models even
in the central rapidity region. Since only a half of the total
luminosity will be utilized for measuring $D_{L\Pi}^{++}$, and the
other half will go to the measurements of $D_{L\Pi}^{+-}$, the
statistical errors for these parameters would be larger than for
$D_{L\Pi}$ by a factor of $\sqrt{2}$. However, it is worth underlining
that $D_{LL}^{++}\approx 2 D_{LL}$ and $D_{LS}^{+-}\approx 2 D_{LS}$
for $\eta$ in the vicinity of zero. These relations follow from
Eq.~(\ref{eq:d_unpol}) with $|A_{LL}|\ll 1$, and taking into account
the ``forward--backward'' symmetry of the initial system of two
colliding protons. As a result, in the central region, the statistical
significance of measurements with two polarized beams would be higher
compared to the case of only one beam being polarized.

\section{Summary}
It is shown that both components, $D_{LL}$ and $D_{LS}$, of the proton
helicity transfer to the polarization of inclusive $\Lambda_{c}^{+}$
hyperons are expected to be equally sensitive to $\Delta G/G$. In the
central region, the expected effects at $\sim$5--15\% are well above
the achievable at RHIC statistical errors, which are also small enough
for distinguishing the {\em AAC}\,\cite{goto:1} and {\em
  GRSV}\,\cite{gluck:1} parameterizations for $\Delta G/G$. The really
large spin transfers at the level of up to 20--30\% are expected at
$\eta\sim 2$ and beyond, which could be potentially accessible at
STAR\cite{star:1}, but definitely with the recently proposed new
RHIC-II detector\cite{harris:1}.

\section*{Acknowledgments}
It is our pleasure to thank
H.~En'yo,
N.~Saito
and
K.~Yazaki
for the useful discussions.

\end{document}